\RequirePackage{ifpdf}
\ifpdf 
\documentclass[pdftex]{sigma}
\else
\documentclass{sigma}
\fi

\begin{document}

\renewcommand{\PaperNumber}{***}

\FirstPageHeading

\ShortArticleName{Singularity and Symmetry Analyses for Tuberculosis
 Epidemics.}

\ArticleName{Singularity and Symmetry Analyses for Tuberculosis
 Epidemics}

\Author{Maba B Matadi~$^\dag$ and Kesh S Govinder~$^\ddag$}

\AuthorNameForHeading{M.B. Matadi and K.S. Govinder}

\Address{$^\dag$~School of Mathematical Sciences, University of Zululand
Private Bag X1001 KwaDlangezwa 3886, Republic of South Africa} 
\EmailD{\href{mailto:email@address}{matadim@unizulu.ac.za}} 

\Address{$^\ddag$~School of Mathematical Sciences, Statistics and Computer Sciences, University of KwaZulu Natal, Republic of South Africa}
\EmailD{\href{mailto:email@address}{govinder@ukzn.ac.za}} 


\ArticleDates{Received ???, in final form ????; Published online ????}

\Abstract{We analyse the model of Tuberculosis due to Blower (\textit{Nature Medecine} \textbf{1(8)} 815-821) from the point
of view of symmetry and singularity analysis. From the study we provide a demonstration 
of the integrability of the model to present an explicit solution.}

\Keywords{Singularity; Symmetry; TB epidemics}

\Classification{?????; ?????; ?????} 
\section{Introduction}
Tuberculosis (TB) is an airborne-transmitted disease and in human
beings is caused by Mycobacterium tuberculosis bacteria (Mtb). Mtb
droplets are released into the air by an infectious individual
coughing and/or sneezing. Tubercle bacillus carried by such
droplets live in the air for a short period of time \cite{Song}
(about 2hours) and therefore it is believed that occasional
contact with an infectious person rarely leads to infection. TB is
described as a slow disease because of its long and variable latency
period and its short and relatively narrow infectious
period \cite{Song}. The initially exposed individuals (infected individuals) have a
higher risk of developing active TB \cite{Blower}. These individuals still face
the possibility of progressing to infectious TB, but the rate of
progression slows. In other words the likelihood of becoming an
active infectious case decreases with the age of the infection.
With this in mind several researchers constructed a series of
dynamical models for TB progression and transmission in scenarios
that took these
factors into consideration \cite{Blower,Blower1,Blower2}.\\

In this paper we analyse the model formulated by Blower \emph{et
al} \cite{Blower} from the point of view of Singularity and Lie analysis. Blower
\emph{et al} divided the population of interest into three
epidemiological classes: susceptible, latent and infectious. The
infection rate given by $\beta{SI}$ (using the law of mass action)
is divided. A portion, $p\beta{SI}$, gives rise to immediately
active cases (fast progression) while the rest, $(1-p)\beta{SI}$,
gives rise to latent-TB cases with a low risk of progressing to
active TB (slow progression) \cite{Blower}. The rate of progression from latent
TB to active TB is assumed to be proportional to the number of
latent-TB cases, that is, it is given by $kE$, where $k$ ranges
from $0.00256$ to $0.00527$ (slow progression) \cite{Blower}. The total
incidence rate is $p\beta{SI}+kE$. In our analysis we focused on
the situation in which the rate of recruitment is equal to the
birth rate and the total population size is constant \cite{Blower}. Thus the
model system is given by \cite{Blower}
\begin{eqnarray}
\frac{dS}{dt}&=&\mu-\beta{S(t)}I(t)-\mu{S(t)},\nonumber\\
\frac{dE}{dt}&=&(1-p)\beta{S(t)}I(t)-kE(t)-\mu{E(t)}\label{eq:slowtb}\\
\frac{dI}{dt}&=&p{\beta{S(t)}}I(t)+kE(t)-\mu{I(t)}.\nonumber
\end{eqnarray}
This paper is organised as followed. In Section 2 we subject the model system (\ref{eq:slowtb}) to the Painlev\'e analysis 
and we present a model as a raw dynamical system which a single second order differential equation. We perform a Lie symmetry 
analysis in Section 3 and obtained the explicit solutions in the case where the infection rate is the sum of the death rate and the
rate of progression from latent TB to active TB. In Section 4 we study in detail and plot the solutions by using 
the link between parameters that the Lie group analysis has given.
\section{Singularity Analysis}
There are four standard approaches to the analysis of nonlinear ordinary
or partial differential equations. The approaches comprise numerical computation, 
dynamical systems analysis, singularity analysis and symmetry analysis, 
all of which possess extensive literatures. Singularity analysis was initiated by Kowalevski \cite{Kowalevski} in her determination of the third integrable
case of the Euler equations for the top and was in large measure developed by the French School developed by Paul Painlev\'e
about the period of La Belle Epoque \cite{Gambier, Gambier1, Chazy}. There have been significant contributions since then.   
For a recent and an erudite contribution to the state of the art see the book edited by Conte \cite{Conte}. For less technical
works devoted to the methodology the interested reader is referred to the text of Tabor \cite{Tabor} and 
the report of Ramani \emph{et al.} \cite{Ramani}. The essence of the singularity analysis of a differential equation is the 
determination of the existence of isolated movable polelite singularities about which one can develop a Laurent expansion 
 containing arbitrary constants equal in number to the order of the system \cite{Nucci}. The location of the singularity is determined 
by the initial conditions of the system. An equation of moderately, or more, complicated structure can possess more than one polelite
singularity \cite{Nucci}.\\

The application of the analysis is usually quite algorithmic \cite{Nucci}. Indeed, it is standard practice to apply 
ARS algorithm \cite{Ablowitz}, although there are instances, of particular relevance to the analysis 
of systems of first-order ordinary linear differential equations typically encountered in the mathematical modelling of epidemics 
\cite{Nucci}, in which the subtler approach advocated by Hua et at. \cite{Hua} 
is to be preferred \cite{Nucci}. The singularity analysis is a poweful tool for construction of symmetries, explicit solutions
and Lie-B\"acklund transformation. It is also helps to find Lax pairs and recursion operators
and plays an important role in the study of a chaotic behaviour of nonlinear differential equations \cite{Mohammad}.
\subsection{Singularity analysis of the three dimensional system} We begin the singularity analysis
of (\ref{eq:slowtb}) in the usual way by substituting
$S=a_{0}\tau^{p_{1}},~E=b_{0}\tau^{p_{2}}~\textrm{and}~I=c_{0}\tau^{p_{3}}$
to determine the leading-order behaviour. We find that
\begin{eqnarray}\label{eq:slowtbn}
p_{1}=p_{2}=p_{3}=-1~\textrm{and}~\left(%
\begin{array}{c}
  a_{0} \\\\\\
  b_{0} \\\\\\
  c_{0} \\
\end{array}%
\right)=\left(%
\begin{array}{c}
  \displaystyle{-\frac{1}{p\beta}} \\\\
  \displaystyle{\frac{1-p}{p\beta}} \\\\
  \displaystyle{\frac{1}{\beta}} \\
\end{array}%
\right).
\end{eqnarray}
The resonances are given by
\begin{equation}
r=-1~\textrm{and}~r=1(2).
\end{equation}
We check for consistency at the resonance by substituting
\begin{equation}
S=a_{i}\tau^{i-1},~E=b_{i}\tau^{i-1}~\textrm{and}~I=c_{i}\tau^{i-1}
\end{equation}
into the full system, (\ref{eq:slowtb}), with the leading-order
term as given in (\ref{eq:slowtbn}). At the resonance $+1$ we
obtain the system
\begin{eqnarray}\label{eq:slowtbf}
\left(%
\begin{array}{ccc}
  1 & 0 & \displaystyle{-\frac{1}{p}} \\\\
  \displaystyle{1-p} & 0 & \displaystyle{-\frac{1-p}{p}} \\\\
  \displaystyle{p} & 0 & -1 \\
\end{array}%
\right)\left(%
\begin{array}{c}
  a_{1} \\\\
  b_{1} \\\\
  c_{1} \\
\end{array}%
\right)=\left(%
\begin{array}{c}
  \displaystyle{\frac{\mu}{p\beta}} \\\\
 \displaystyle{\frac{(1-p)(k+\mu)}{p\beta}}\\\\
  \displaystyle{\frac{\mu}{\beta}-\frac{k(1-p)}{p\beta}} \\
\end{array}%
\right)
\end{eqnarray}
The three equations in (\ref{eq:slowtbf}) are identical if and
only if $k=0$. The system (\ref{eq:slowtb}) is consistent if the transition rate from the exposed class to the
infectious class is zero. Therefore (\ref{eq:slowtb}) passes the
Painlev\'e test and is integrable in the sense of Poincar\'e for the constraint $k=0$.
\subsection{Singularity analysis of the two dimensional system}
By assuming that the total population size is constant, we have
\begin{equation}\label{eq:slowtb1}
N(t)=S(t)+E(t)+I(t)=1\equiv{\textrm{Constant}}.
\end{equation}
We derive $I(t)$ from (\ref{eq:slowtb1}), i.e.
\begin{equation}\label{eq:slowtb2}
I(t)=1-S(t)-E(t),
\end{equation}
the three-dimensional system (\ref{eq:slowtb}) is reduced
to the two-dimensional system
\begin{eqnarray}
\dot{S}&=&\mu+\beta{S^{2}}+\beta{SE}-(\beta+\mu){S}\label{eq:slowtb3}\\
\dot{E}&=&(1-p)\beta{S}-(1-p)\beta{S^{2}}-(1-p)\beta{SE}-(k+\mu)E\label{eq:slowtb4}.
\end{eqnarray}
The exponents for the usual leading-order behaviour substitution,
$S=\alpha_{1}\tau^{q_{1}}~\textrm{and}~E=\alpha_{2}\tau^{q_{2}}$ are\\
$q_{1}-1$~~:~~$q_{1}$~~$2q_{1}$~~$q_{1}+q_{2}$\\
$q_{2}-1$~~:~~$q_{1}$~~$2q_{1}$~~$q_{1}+q_{2}$~~$q_{2}$\\

The case $q_{1}=q_{2}=-1$ contains the left-hand side and the second
and the third terms of the right-hand side of (\ref{eq:slowtb3}) and the 
left-hand side and the second and the third terms of the right-hand side of (\ref{eq:slowtb4})
as dominant terms.\\

The coefficients of the leading-order terms are
\begin{eqnarray}\label{eq:maba}
\alpha_{1}=\frac{p-2}{\beta}~~~\alpha_{2}=-\frac{p-1}{p\beta}
\end{eqnarray}

In the case that the leading-order behaviour does not provide the correct number of arbitrary 
constants the generic situation illustrated by the results given in (\ref{eq:maba}) it is
necessary to determine whether there exists a term, or terms, at which the requisite number of 
arbitrary constants can enter. The powers at which these arbitrary constants enter are almost 
called resonances on occasion they are also known as Kowalevski exponents (after the pioneering
woman in this area). To determine the resonances we substitute
\begin{eqnarray}\label{eq:matadi}
S=\alpha_{1}+m\tau^{r-1}~~~E=\alpha_{2}+n\tau^{r-1}
\end{eqnarray}
into (\ref{eq:slowtb3}) and (\ref{eq:slowtb4}) and collect the linear terms in $m$ and $n$ we obtain
\begin{eqnarray}\label{eq:slowt6}
\left(%
\begin{array}{cc}
  r-p+2 & -p+2 \\\\
  -(p-1)(p-3)& r-p^{2}+3p-3  \\\\
 \end{array}%
\right)\left(%
\begin{array}{c}
  m \\\\
 n \\\\
  \end{array}%
\right)=0
\end{eqnarray}
The requirement that the system be consistent leads to the equation
\begin{equation}\label{eq:boni}
 r^{2}-r(p-1)^{2}-(2-p)p=0.
\end{equation}
For $p=1$ the solution of (\ref{eq:boni}) is $r=\pm{1}$.\\

The second approach to the determination of the resonances and the question of consistency
mentioned in \cite{Hua} is particularly suited to a system containing a selection of parameters. Typically
such systems are only integrable subject to some constraint(s) on the parameters. After the nature 
of the polelike singularity is identified, we subtitute
\begin{equation}
 S=\sum_{j=0}^{\infty}a_{i}\tau^{i-1},~E=\sum_{j=0}^{\infty}b_{i}\tau^{i-1}
\end{equation}
to obtain
$$\sum_{i=0}^{\infty}\{(i-1)a_{i}\tau^{i-2}+(\beta+\mu)a_{i}\tau^{i-1}-\sum_{j=0}^{\infty}[\beta(a_{i}a_{j}+a_{i}b_{j})]\tau^{i+j-2}\}-\mu=0$$
$$\sum_{i=0}^{\infty}\{(i-1)b_{i}\tau^{i-2}-[(1-p)\beta{a_{i}-(k+\mu)b_{i}}]\tau^{i-1}+\sum_{j=0}^{\infty}[(1-p)\beta(a_{i}a_{j}+a_{i}b_{j})]\tau^{i+j-2}\}=0$$
We illustrate the workings of the algorithm \cite{Geronimi} with the first few powers.\\

$\tau^{-2}:$
\begin{eqnarray}\label{eq:boni1}
-a_{0}-\beta(a_{0}^{2}+a_{0}b_{0})&=&0\label{eq:boni1}\\
-b_{0}+(1-p)\beta(a_{0}^{2}+a_{0}b_{0})&=&0\label{eq:boni2}
\end{eqnarray}
from (\ref{eq:boni1}) and (\ref{eq:boni2}) we have 
\begin{eqnarray}
a_{0}=\frac{p-2}{\beta},~~~b_{0}=-\frac{p-1}{p\beta}: \textrm{not a resonance, then we continue}\nonumber
\end{eqnarray}
and 
\begin{eqnarray}
b_{0}=(1-p)a_{0}: \textrm{a resonance, then we stop}\nonumber
\end{eqnarray}
$\tau^{-1}:$
\begin{equation}\nonumber
\left(%
\begin{array}{cc}
  2\beta{a_{0}}+\beta{b_{0}} & \beta{a_{0}} \\\\
 2(1-p)\beta{a_{0}}+(1-p)\beta{b_{0}}& (1-p)\beta{a_{0}}\\\\
\end{array}%
\right)\left(%
\begin{array}{c}
  a_{1} \\\\
b_{1}\\\\
\end{array}%
\right)=\left(%
\begin{array}{c}
  (\beta+\mu)a_{0} \\\\
(1-p)\beta{a_{0}}-(k+\mu)b_{0}\\\\
\end{array}%
\right)
\end{equation}
we have  $a_{1}=b_{1}=0: \textrm{not a resonance, then we continue}$\\
and we have a resonance if
\begin{equation}\label{eq:boni3}
\beta=\frac{(k+\mu)(p-1)}{p(p-2)}+\frac{k+\mu}{p}.
\end{equation}
$\tau^{0}:$
\begin{equation}\nonumber
\left(%
\begin{array}{cc}
  1-2\beta{a_{0}}-\beta{b_{0}} & -\beta{a_{0}} \\\\
 2(1-p)\beta{a_{0}}+(1-p)\beta{b_{0}}& 1+(1-p)\beta{a_{0}}\\\\
\end{array}%
\right)\left(%
\begin{array}{c}
  a_{2} \\\\
b_{2}\\\\
\end{array}%
\right)=\left(%
\begin{array}{c}
  -(\beta+\mu)a_{1} \\\\
(1-p)\beta{a_{1}}-(k+\mu)b_{1}\\\\
\end{array}%
\right)
\end{equation}
recall $a_{1}=b_{1}=0$, we have a resonance if
\begin{equation}\label{eq:tola}
 p^{2}-2p+1=0
\end{equation}
There are two identical real root of (\ref{eq:tola}). Equations (\ref{eq:slowtb3}) and (\ref{eq:slowtb4})
possesses the Painlev\'e property for $p=1$.
\subsection{Singularity analysis of the second-order ordinary differential equation}
From (\ref{eq:slowtb3}) we have
\begin{equation}\label{eq:slowtb5}
E=\frac{\dot{S}-\mu}{\beta{S}}-S+\frac{(\mu+\beta)}{\beta}.
\end{equation}
The derivative of (\ref{eq:slowtb5}) gives
\begin{equation}\label{eq:slowtb6}
\dot{E}=\frac{\ddot{S}}{\beta{S}}-\frac{\dot{S}^{2}}{\beta{S^{2}}}+\frac{\mu\dot{S}}{\beta{S^{2}}}-\dot{S}.
\end{equation}
The substitution of (\ref{eq:slowtb5}) and (\ref{eq:slowtb6}) into
(\ref{eq:slowtb4}) gives the following second-order equation for
$S(t)$
\begin{eqnarray}\label{eq:slowtb7}
&&S\ddot{S}-\dot{S}^{2}-p\beta\dot{S}S^{2}+(k+\mu)\dot{S}S+\mu\dot{S}-\beta{(k+p\mu)}S^{3}-\mu{(k+\mu)S}\nonumber\\
&&-[\beta\mu(1-p)-(k+\mu)(\mu+\beta)]S^{2}=0
\end{eqnarray}
The first three terms in (\ref{eq:slowtb7}) are dominant, the exponent of the leading
order term is $-1$ and the resonances are at $\pm{1}$. To establish that there is
consistency for the exponent at which the resonance occurs, we substitute
\begin{eqnarray}\label{eq:slowtb8}
S=a_{i}\tau^{i-1}
\end{eqnarray}
into (\ref{eq:slowtb7}) to obtain
\begin{eqnarray}\label{eq:slowtb9}
&&(i-1)(i-2)a_{i}a_{j}\tau^{i+j-4}-(i-1)(j-1)a_{i}a_{j}\tau^{i+j-4}-(i-1)p\beta{a_{i}a_{j}a_{k}\tau^{i+j+k-4}}\nonumber\\
&&+(i-1)(k+\mu)a_{i}a_{j}\tau^{i+j-3}+(i-1)\mu{a_{i}\tau^{i-2}}-\beta{(k+p\mu)}a_{i}a_{j}a_{k}\tau^{i+j+k-3}\nonumber\\
&&-\mu{(k+\mu)S}a_{i}\tau^{i-1}-[\beta\mu(1-p)-(k+\mu)(\mu+\beta)]a_{i}a_{j}\tau^{i+j-2}=0\nonumber
\end{eqnarray}
The coefficients of $\tau^{-4}$ gives
\begin{equation}\label{eq:slowtb10}
 2a_{0}^{2}-a_{0}^{2}+p\beta{a_{0}^{3}}\Longrightarrow{a_{0}=\frac{1}{p\beta}}
\end{equation}
and the coefficients of $\tau^{-3}$ gives
\begin{equation}\label{eq:slowtb11}
 2a_{0}a_{1}-p\beta{a_{0}(2a_{0}a_{1})}-(k+\mu)a_{0}^{2}-\beta(k+p\mu)a_{0}^{3}=0.
\end{equation}

From the result in (\ref{eq:slowtb10}) the coefficient of $a_{1}$ is zero as is to be expected
as this is where the resonance occurs. The terms remaining in (\ref{eq:slowtb11}) give the condition
\begin{equation}\label{eq:slowtb12}
 k=\frac{-2p\mu}{p+1}
\end{equation}
Subject to the constraint (\ref{eq:slowtb12}), the second order ordinary differential equation (\ref{eq:slowtb7})
has an analytic solution for $S(t)$. It follows from (\ref{eq:slowtb5}) that $E(t)$ is also analytic and from
(\ref{eq:slowtb2}) that $I(t)$ is also analytic.\\

For $p=1$ the condition (\ref{eq:slowtb12}) becomes
\begin{equation}\label{eq:slowtb13}
 k+\mu=0
\end{equation}
Condition (\ref{eq:slowtb13}) means that if $p=1$ then the sum of the tansition rate from exposed class to infectious class
and the natural death rate is zero.
\section{Lie Analysis}
An $n$th order ordinary differential equation 
\begin{equation}
N(x,y,y',\ldots,y^{(n)}) = 0  \label{l1}
\end{equation} admits the
one parameter Lie group of transformations 
\begin{eqnarray}
\bar{x} &=& x + \varepsilon \xi \\
\bar{y} &=& y + \varepsilon \eta 
\end{eqnarray} with infinitesimal generator
\begin{equation}
 G = \xi\frac{\partial}{\partial{x}} + \eta\frac{\partial}{\partial{y}} \label{l2} 
\end{equation} 
 if \begin{equation}
 G^{[n]}
N_{\left|_{_{N=0}}\right.} = 0, \label{l3} 
\end{equation} where $G^{[n]}$ is
the $n$ th extension of $G$
 given by  
\begin{equation} 
G^{[n]} = G +\sum_{i=1}^n \left\{ \eta^{(i)} - \sum_{j=0}^{i-1}
\left(\begin{array}{c} i \\  j \\ 
\end{array} \right) y^{(j+1)}
\xi^{(i-j)} \right\}\frac{\partial}{\partial{y^{(i)}}}. \label{l4} 
\end{equation} We say that the equation possesses the
symmetry (group generator) 
\begin{equation}
 G = \xi {\partial{x}} + \eta{\partial{y}} 
\end{equation}  iff (\ref{l3}) holds.\\

The autonomous system (\ref{eq:slowtb}) possesses the obvious Lie point symmetry $\partial_{t}$. 
The integrability from the point of view of Lie requires the knowledge of a three-dimensional solvable algebra. 
We acknowledge that system (\ref{eq:slowtb}) is a first-order ordinary differential equations and possesses an infinite 
number of Lie point symmetries. Since infinity is not a satisfactory number, we apply Lie group analysis to the second order 
ordinary differential equation (\ref{eq:slowtb7}) and obtain non-trivial Lie point symmetries in two special cases:
$$\textrm{case (1)}$$
$$\beta+\mu=0, k\neq{0}$$
In this case we obtain an eight-dimensional Lie symmetry algebra, namely 
\begin{eqnarray}
G_{1}&=&-\frac{\exp{[-kt]}}{k}S\partial_{S}\nonumber\\
G_{2}&=&S\partial_{S}\nonumber\\
G_{3}&=&\frac{\exp{[-kt]}}{k^{2}}\partial_{t}-\frac{\exp{[-kt]}}{k}S\log{[S]}\partial_{S}\nonumber\\
G_{4}&=&S\exp{[S]}\partial_{S}\nonumber\\
G_{5}&=&\frac{\exp{[kt]}}{k}\log{S}\partial_{t}\nonumber\\
G_{6}&=&\log{[S]}\partial_{t}-kS\log{[S]}^{2}\partial_{S}\nonumber\\
G_{7}&=&\frac{\exp{[kt]}}{k}\partial_{t}\nonumber\\
G_{8}&=&\partial_{t}\nonumber
\end{eqnarray}
The above Lie symmetry algebra is isomorphic to $Sl(3,\Re)$. This means that equation (\ref{eq:slowtb7}) 
is linearizable by means of a point transformation.
 $$\textrm{case (2)}$$
 $$\beta\neq{k+\mu},~p\neq{1}$$
This case provides the coefficients of the infinitesimal generator below:
\begin{eqnarray}
\xi{(t,S)}&=&C_{1}+\exp{[(\beta{-k}-\mu+p)t]}C_{2}+\frac{\exp{[{t}]}C_{3}}{S}+\exp{[(\beta{-k}-\mu+p)t]}S^{2}C_{4}\nonumber\\
&&+\exp{[{t}]}S^{2}C_{5}+SC_{6}-\exp{[(\beta-k-\mu+p)t]}(C_{7}-C_{8})\label{eq:modelHIV7}\\
\eta{(t,S)}&=&-{\exp{[(\beta-k-\mu+p)t]}SC_{2}}-\frac{(\beta-k-\mu+p)\exp{[-\mu{t}]}C_{3}}{S}\nonumber\\
&&+\exp{[(\beta\mu-p)t]}S\Big[(\beta\mu-p)C_{7}-\mu{C_{8}}\Big]\label{eq:modelHIV8}
\end{eqnarray}
which mean that, if $\beta\neq{k+\mu},~p\neq{1},$ the
eight-dimensional Lie symmetry algebra is generated by the following
eight operators:
\begin{eqnarray}
G_{1}&=&\partial_{t},\nonumber\\
G_{2}&=&\exp{[(\beta-k-\mu)t]}\partial_{t}-\exp{[(\beta{-k}-\mu+p)t]}S\partial_{S},\nonumber\\
G_{3}&=&\frac{1}{S}\exp{[t]}\partial_{t}-\frac{(\beta-k-\mu+p)\exp{[-\mu{t}]}}{S}\partial_{S},\nonumber\\
G_{4}&=&\exp{[\beta-k-\mu{t}+p]}S^{2}\partial_{t},\\
G_{5}&=&\exp{[t]}S^{2}\partial_{t},\nonumber\\
G_{6}&=&S\partial_{t},\nonumber\\
G_{7}&=&-\exp{[(\beta-k-\mu+p)t]}\partial_{t}+(\beta\mu-p)\exp{[(\beta\mu-p)t]}S\partial_{S},\nonumber\\
G_{8}&=&\exp{[(\beta-k-\mu+p)t]}\partial_{t}-\mu{S}\exp{[\beta\mu-p]t}\partial_{S}.\nonumber
\end{eqnarray}
In our analysis we assume that $p=1$. For the case $p\neq{1}$ the result is given by the numerical
simulations \cite{Blower}.  However, if $\beta=k+\mu$ (see equation (\ref{eq:boni3}) if $p=1$), 
then the eight Lie symmetries are
\begin{eqnarray}
G_{1}&=&\partial_{t},\nonumber\\
G_{2}&=&\exp{[t]}[\partial_{t}-S\partial_{S}],\nonumber\\
G_{3}&=&\frac{1}{S}\exp{[t]}\partial_{t}-\frac{\exp{[-\mu{t}]}}{S}\partial_{S},\nonumber\\
G_{4}&=&S^{2}\partial_{t},\\
G_{5}&=&\exp{[t]}S^{2}\partial_{t},\nonumber\\
G_{6}&=&S\partial_{t},\nonumber\\
G_{7}&=&-\partial_{t}+[(k+\mu)\mu-p]\exp{[(k+\mu)\mu-1)t]}S\partial_{S},\nonumber\\
G_{8}&=&\partial_{t}-\mu{S}\exp{[k+\mu-1]t}\partial_{S}.\nonumber
\end{eqnarray}
We note that $[G_{1},~G_{2}]_{LB}=G_{2}$ gives a 
reduction of (\ref{eq:slowtb7}) by
\begin{eqnarray}\label{eq:win12}
G_{2}&=&e^{t}(\partial_{t}-S(t))\partial_{S}.
\end{eqnarray}
The associated Lagrange's system for the zeroth and first-order
invariant of $G_{2}$ in (\ref{eq:slowtb7}) is
$$\frac{dt}{1}=\frac{dS}{-S}=\frac{d\dot{S}}{-2\dot{S}-S}$$
so that
\begin{equation}\label{eq:tola1}
T=t+\log{S},~~~~~U=\frac{\dot{S}}{S^{2}}+\frac{1}{S}
\end{equation}
with $U$  and $T$ the new dependent and independent variables,
respectively. Therefore equation (\ref{eq:slowtb7})
becomes
$$\frac{dU}{dT}+U+1=0$$
which can be easily integrated to give
\begin{equation}\label{eq:tola2}
(U+1)\exp[T]=A.
\end{equation}
Equation (\ref{eq:tola2}) becomes
\begin{equation}\label{eq:tola3}
(U+1)=A\exp[-T].
\end{equation}
The substitution of (\ref{eq:tola1}) into (\ref{eq:tola3}) gives
\begin{equation}\label{eq:tola4}
\frac{\dot{S}}{S}+S+1=A\exp[-t].
\end{equation}
We integrate (\ref{eq:tola4}) to obtain
\begin{equation}\label{eq:tola5}
S(t)=\frac{AB\exp[-t]}{D\exp[{A\exp[-t]}]+B}.
\end{equation}
The derivative of (\ref{eq:tola4}) gives
\begin{eqnarray}\label{eq:tola8}
\dot{S}(t)&=&-\frac{AB\exp[-t]}{[D\exp[{A\exp[-t]}]+B]}\nonumber\\
&-&\frac{A^{2}BD\exp[-2t]\exp[{A\exp[-t]}]}{[D\exp[{A\exp[-t]}]+B]^{2}}.
\end{eqnarray}
Substituting (\ref{eq:tola5}) and (\ref{eq:tola8}) into
(\ref{eq:slowtb5}) we have
\begin{eqnarray}\label{eq:tola9}
&&E(t)=\frac{(\beta+\mu)}{\beta}-\frac{\mu[D\exp[{A\exp[-t]}]+B]}{AB\beta\exp[-t]}\nonumber\\
&-&\frac{A\exp[-t][D\exp[{A\exp[-t]}]+B]+B]}{\beta[D\exp[{A\exp[-t]}]+B]}.
\end{eqnarray}
The substitution of (\ref{eq:tola5}) and (\ref{eq:tola9}) into
(\ref{eq:slowtb2}) gives
\begin{eqnarray}\label{eq:tola10}
I(t)&=&1-\frac{AB\exp[-t]}{D\beta\exp[{A\exp[-t]}]+B\beta}+\frac{\mu[D\exp[{A\exp[-t]}]+B]}{AB\exp[-t]}\nonumber\\
&+&\frac{A\exp[-t][D\exp[{A\exp[-t]}]+B]+B]}{\beta[D\exp[{A\exp[-t]}]+B]}\nonumber\\
&+&\frac{AB+(\beta+\mu)}{\beta}.\nonumber
\end{eqnarray}
The case $k+\mu=0$ (see equation (\ref{eq:slowtb13}) if $p=1$) reduces equation (\ref{eq:slowtb7}) to
\begin{equation}\label{eq:slowtb14}
 S\ddot{S}-\dot{S}^{2}-\beta\dot{S}S^{2}+\mu\dot{S}=0
\end{equation}
The only condition for equation (\ref{eq:slowtb14}) to pass the Painlev\'e test is when $\mu=0$ \cite{Leachp} and from (\ref{eq:slowtb13})
we have $k=0$ as it did for the system of three first order differential equations (\ref{eq:slowtb}). A standard
method for treating a nonlinear equation such as (\ref{eq:slowtb14}) is to raise it to a higher order by 
means of the following Ricatti transformation \cite{Leachp}
\begin{equation}\label{eq:slowtb15}
 S=-\frac{1}{\beta}\frac{\dot{\omega}}{\omega}
\end{equation}
When the transformation (\ref{eq:slowtb15}) is applied to (\ref{eq:slowtb14}), we obtain
\begin{eqnarray}\label{eq:slowtb16}
\frac{1}{\beta^{2}}\Big(\frac{\dddot{\omega}\dot{\omega}-\ddot{\omega}^{2}+\mu\ddot{\omega}\dot{\omega}}{\omega^{2}}\Big)-\frac{\mu}{\beta^{2}}\frac{\dot{\omega}^{3}}{\omega^{3}}=0
\end{eqnarray}
with $\mu=0$ we obtain a generalised Kummer-Schwaz equation \cite{Leachp}
\begin{eqnarray}\label{eq:slowtb17}
\dddot{\omega}\dot{\omega}-\ddot{\omega}^{2}=0
\end{eqnarray}
Equation (\ref{eq:slowtb17}) has three Lie point symmetries \cite{Leachp}, namely
\begin{eqnarray}\label{eq:slowtb18}
&&G_{1}=\partial_{t},\nonumber\\
&&G_{2}=\partial_{\omega}\\
&&G_{3}=\omega\partial_{\omega}\nonumber
\end{eqnarray}
The third symmetry $G_{3}$ is unexpected \cite{Leach and Kesh} and arises since the symmetry associated
with the Ricatti transformation is not the normal subgroup of the two symmetries $\partial_{\omega}$ and 
$\omega\partial_{\omega}$. One reduces (\ref{eq:slowtb17}) to (\ref{eq:slowtb14}) using $\omega\partial_{\omega}$, 
$\partial_{\omega}$ which ceases to be a point symmetry \cite{Leachp}. Rather it becomes the exponential nonlocal symmetry \cite{Geronimi}
$S\exp{[\beta{Sdt}]}\partial_{S}$ and $\partial_{\omega}$ is a Type I hidden symmetry.\cite{Abraham}\\

We reduce equation (\ref{eq:slowtb17}) to a second-order ordinary differential equation by choosing
the operator $G_{2}$. The variables for the reduction are
\begin{equation}\label{eq:slowtb19}
 T=t,~~~W=\log{\dot{\omega}}
\end{equation}
we obtain the linear second order order differential equation
\begin{equation}\label{eq:slowtb20}
 \ddot{W}=0
\end{equation}
Equation (\ref{eq:slowtb20}) has eight Lie point symmetries with the algebra $sl(3,R)$. The solution of (\ref{eq:slowtb20}) is
\begin{equation}\label{eq:slowtb21}
 W(T)=AT+B
\end{equation}
using the change of variables (\ref{eq:slowtb19}) we obtain
\begin{equation}\label{eq:slowtb22}
 \omega(t)=B\int{\exp[At]dt}+C
\end{equation}
therefore from (\ref{eq:slowtb15}) we have
\begin{equation}\label{eq:tola19}
 S(t)=-\frac{1}{\beta}\frac{\exp[At]}{\int{\exp[At]dt}+B}
\end{equation}
Hence, from (\ref{eq:slowtb2}) and (\ref{eq:slowtb5}) we have
\begin{eqnarray}\label{eq:tola19}
&& E(t)=\frac{\beta+\mu}{\beta}-\frac{-A(1+Be^{-t})\exp[-t+Be^{-t}]}{A\beta{\exp[-t+Be^{-t}]}}-\frac{A\mu\exp[-t+Be^{-t}]}{\beta(A\int\exp[-t+Be^{-t}]dt+C)}\nonumber\\
&&-\frac{-A(\exp[-t+Be^{-t}])^{2}}{\beta\exp[-t+Be^{-t}]\times{(A\int\exp[-t+Be^{-t}]dt+C)}}-\frac{A\exp[-t+Be^{-t}]}{(A\int\exp[-t+Be^{-t}]dt+C)}\nonumber
\end{eqnarray}
and
\begin{eqnarray}\label{eq:tola19}
&&I(t)=1-\frac{A\exp[-t+Be^{-t}]}{A\int{\exp[-t+Be^{-t}]dt}+C}-\frac{\beta+\mu}{\beta}+\frac{-A(1+Be^{-t})\exp[-t+Be^{-t}]}{A\beta{\exp[-t+Be^{-t}]}}\nonumber\\
&&+\frac{A\mu\exp[-t+Be^{-t}]}{\beta(A\int\exp[-t+Be^{-t}]dt+C)}+\frac{A\exp[-t+Be^{-t}]}{(A\int\exp[-t+Be^{-t}]dt+C)}\nonumber\\
&&+\frac{-A(\exp[-t+Be^{-t}])^{2}}{\beta\exp[-t+Be^{-t}]\times{(A\int\exp[-t+Be^{-t}]dt+C)}}\nonumber
\end{eqnarray}
\section{Discussion}
In this Section we study in detail the solutions in closed form that we have obtained in the cases $\beta=k+\mu$ 
and $k+\mu=0$. We plot our solutions with the help of the graphing capability of MATHEMATICA. Note that the qualitative
description by Blower \emph{et al} is limited to the study of equilibrium points, their stability and bifurcation diagrams. 
We use the same numerical values of $k$ and $\mu$ as given in \cite{Blower}. The numerical value of $\beta$ is derived from 
the relationship that Lie group analysis has determined.\\

In all the figures, the red line represents the plot of $S$, the blue line represents the plot of $E$ and 
the green line represents the plot of $I$.\\

In the case where the infection rate is the sum of the death rate and the rate of progression from latent TB to active TB. 
We simulate the dynamics of the model by assuming three different values of $k$ within the range considered in \cite{Blower}. In
Figure 1 and Figure 2 we show the dynamics of the model if $\beta=0.50357$ and $\beta=0.50527$ respectively.
\begin{figure}[!ht]
        \centering
        \includegraphics[scale=0.733]{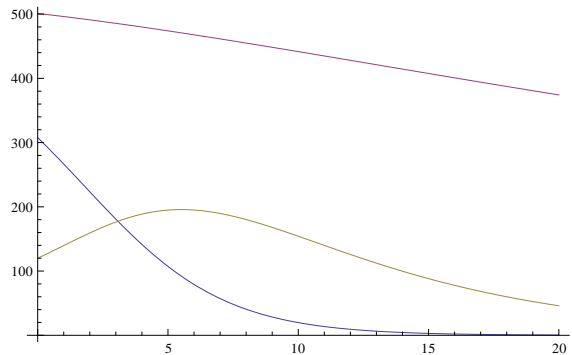}
        \caption{$\beta=0.50357$, $k=0.00357$, $\mu=0.5$.}
        \label{fig}
\end{figure}
\begin{figure}[!ht]
        \centering
        \includegraphics[scale=0.733]{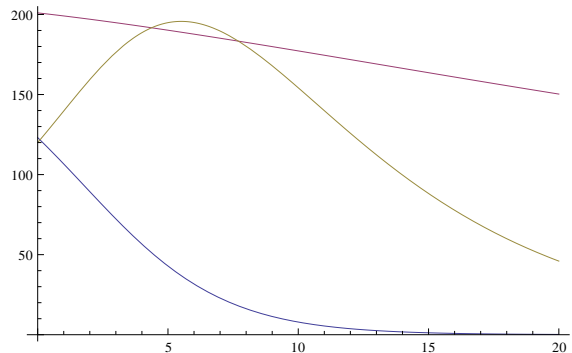}
        \caption{$\beta=0.50527$, $k=0.00527$, $\mu=0.5$.}
        \label{fig}
\end{figure}
\section{Conclusion}
We applied the method of Lie Symmetry and Singularity analysis to
a Mathematical Model which describes Tuberculosis
infection. Lie group analysis is
indeed the most powerful tool to find the general solution of
ordinary differential equations. The Singularity analysis of the model 
reveals that at the resonance $r=\pm{1}$, the parameter $p$ must be one for 
the system to pass the Painlev\'e test. Lie symmetry analysis allowed us to integrate the TB
model by quadrature and we found the general solution of the
model if the infections rate of TB patients is the sum of the death
rate of TB infecteds plus the rate of progression from latent to active TB.

\LastPageEnding

\end{document}